\documentclass[twocolumn]{aastex701}

\def \aap  {A\&A}

\def \apjs  {ApJS}
\def \apjl  {ApJL}

\newcommand{\thbn}{\theta_{Bn}}
\usepackage{subfigure}
\usepackage{graphicx}  
\usepackage{xcolor}
\newcommand{\es}[1]{\textcolor{blue}{[ES: #1]}}
\newcommand{\dc}[1]{\textcolor{magenta}{[DC: #1]}}

\begin{document}

\title{A ``Faux-Shock" Method for Hybrid Simulations of Astrophysical Shocks}

\author[0000-0002-9386-630X]{Emily Simon}
\affiliation{Department of Astronomy and Astrophysics, University of Chicago, 5640 S Ellis Ave, Chicago, IL 60637, USA}
\email[show]{ersimon@uchicago.edu}

\author[0000-0003-0939-8775]{Damiano Caprioli}
\affiliation{Department of Astronomy and Astrophysics, University of Chicago, 5640 S Ellis Ave, Chicago, IL 60637, USA}
\affiliation{Enrico Fermi Institute, The University of Chicago, Chicago, IL 60637, USA}
\email{ersimon@uchicago.edu}

\author[0000-0002-2160-7288]{Colby C. Haggerty}
\affiliation{Institute for Astronomy, University of Hawaii, Honolulu, HI 96822, USA}
\email{ersimon@uchicago.edu}

\author[0000-0002-3778-1432]{Brian Reville}
\affiliation{Max-Planck-Institut f\"ur Kernphysik,
Saupfercheckweg 1, Heidelberg, 69117, Germany}
\email{ersimon@uchicago.edu}



\begin{abstract}
We demonstrate a novel setup for hybrid particle-in-cell simulations designed to isolate the physics of the shock precursor over long time periods for significantly lower computational cost than previous methods.
This is achieved using a ``faux-shock" or shock-like boundary condition on one edge of our simulation domain such that particles that interact with the boundary either pass through it or are reflected off of it with a change in momentum that mimics scattering in the downstream. 
We show that our faux-shock setup reproduces the same fluid quantities and phase spaces as traditional shock simulations, including those which could otherwise only be done in 3D, with higher particle resolution and for reduced computational cost.
While the method involves an assumed boundary condition, it nonetheless captures the essential physics of interest, establishing it as a reliable and efficient tool for future self-consistent studies of instabilities driven by cosmic rays in a shock upstream medium.

\end{abstract}

\section{Introduction}
Diffusive shock acceleration (DSA) occurring at astrophysical shocks  \citep{krymskii77,axford+77p, bell78a, blandford+78} is thought to be responsible for the acceleration of galactic cosmic rays (CRs) up to $10^{15}$ eV and possibly beyond \citep[e.g.,][]{bell+13, cristofari+22, vieu+22}.
DSA occurs across a wide range of astrophysical environments, from heliospheric shocks \citep[e.g.,][]{guo+13,johlander+16a,turner+18} and supernova remnants (SNRs) \citep[e.g.,][]{lagage+83a, hillas05, morlino+12} to galaxy clusters \citep[e.g.,][]{brunetti+14,vazza+17,boess+24}, making it a universal process wherever shocks are present. 

Although DSA is widely invoked in models of CR acceleration, our understanding of the precise mechanisms by which CRs are accelerated at their respective sources and to what maximum energies is still incomplete. 
These uncertainties are largely due to the ways in which CRs modify shock hydrodynamics 
and drive the growth of plasma instabilities \cite[e.g.][]{drury83, bell04, haggerty+24}.
In particular, it was pointed out that the non-resonant CR streaming instability \citep{bell04}, also referred to as the Bell instability, could lead to nonlinear magnetic field amplification and subsequently higher CR maximum energies \citep[e.g.,][]{bell+13}. 

Simulations are the most effective tool for studying this process due to its nonlinear nature.
In particular, kinetic simulations are required to capture the feedback of the instability on CRs themselves.
However, self-consistently simulating this regime is computationally challenging.
The instability occurs in the shock upstream and must begin growing sufficiently far away from the shock for the magnetic field to reach the nonlinear regime before advection sweeps the fields into the downstream.
In most kinetic simulations, a shock is formed when a fluid hits a reflective wall and causes the reflected flow and the incoming flow to collide \citep[e.g.,][]{winske+88}. 
This setup means that, for a fixed box, the shock propagates through the simulation domain, gradually shrinking the upstream length in time, thus limiting the available space for instabilities to grow in. 
Similarly, the transverse box size in $2$ or $3$ dimensions must be sufficiently large to allow the filamentation and cavities typical of Bell to grow without becoming truncated or only partially resolved \citep{reville+12, caprioli+13}.
For typical astrophysical parameters resembling a SNR, the computational burden quickly becomes unfeasible;
resolving the ion skin depth ($d_{\rm i} \sim 10^5$ cm) and plasma timescales (seconds) is essential for capturing the relevant microphysics, yet SNRs span many parsecs and evolve over tens of thousand years.
Even in 1D, bridging this wide range of scales is computationally intractable. 

To address these computational challenges, we have devised a novel setup for hybrid PIC simulations designed to isolate the physics of the shock precursor while significantly reducing computational cost.
This enables longer simulations in larger domains than previously feasible, providing a controlled platform for studying CR-driven magnetic field amplification and particle acceleration with near self-consistency.
This is achieved by the implementation of a custom boundary condition on one wall of the simulation domain which we call a ``\textit{faux-shock}".
The faux-shock (hereafter: FS) mimics the physics of a shock, allowing some particles to pass and reflecting others with an energy kick as if they've been scattered in the downstream. 
Because the FS sits permanently on one wall of the box, the simulation is fixed in the shock frame, meaning the upstream has a constant length for the entirety of the simulation.

The trade-off of this method is giving up the ability to self-consistently account for all the effects that occur at the shock, such as particle injection, as well as downstream particle scattering on the post-shock magnetic turbulence, which controls the CR anisotropies and drift motions \citep[e.g.,][]{bell+11,haggerty+20, caprioli+20} which can affect the CR spectrum. 
Nevertheless, our formalism is flexible enough to allow us to implement all of these modifications by prescribing the rate of return of CRs from the downstream, their anisotropy, and their energy gain, calibrated via global hybrid and PIC simulations.
The advantage of the presented formalism is twofold: on the one hand, we introduce CRs as a separate population from the thermal ions, which allows us to increase the statistics of energetic particles at modest computational cost and, on the other hand, this allows us to run shock-frame simulations in unprecedentedly-large boxes and for times long enough to observe the long-term evolution of the magnetic field amplification.

The paper is organized as follows: in \S\ref{section:setup} we discuss the specifics of the FS simulation setup and how it differs from traditional reflecting wall shock simulations. In \S\ref{sec:results} we show results from the FS setup and compare them to benchmark reflecting wall shock simulations, and we conclude in \S\ref{sec:conclusions}.
Practical applications of the presented formalism will be presented in forthcoming works.

\section{Faux-Shock Simulation Setup}
\label{section:setup}

\begin{table*}
\centering
\begin{tabular}{|c|c|c|c|c|c|c|c|c|c|c|}
    \hline
    Run & Setup & $M_{\rm A}$ & $c/v_{\rm A}$ & $L_x/d_{\rm i}$ & $L_\perp / d_{\rm i}$ & $\Delta x / d_{\rm i}$ & $\Delta t \Omega_{ci}$ & $\thbn$ & $n_{\rm cr}/n_{\rm g}$ & $p_{\rm iso}/{mv_{\rm A}}$\\
    \hline
    $\mathcal{A}$ & faux-shock & $20$ & $1000$ & $10000$ & $300$ & 1.5 & 0.005 & 0 & 0.01 & 100 \\
    $\mathcal{B}$ & faux-shock & $100$ & $10000$ & $10000$ & $50$ & 1.5 & 0.00005 & 80 & 0.01 & 200 \\
    \hline
    $\mathcal{C}$ & Reflecting Wall & $20$ & $1000$ & $20000$ & $300$ & 1.5 & 0.005 & 0 & N/A & N/A \\
    $3\mathcal{D}$ & Reflecting Wall & $20$ & N/A & $10000$ & $50$ & 2.5 & 0.00075 & 80 & N/A & N/A \\
    \hline
\end{tabular} 
\caption{List of the simulations considered in this work. Runs A, B, and C are performed in 2D with \texttt{dHybridR}. Run $3\mathcal{D}$ is performed in 3D with \texttt{dHybrid}, a strictly non-relativistic version of the same code. The speed of light in Run B is set sufficiently high to ensure CRs remain non-relativistic for accurate comparison to Run $3\mathcal{D}$ . Runs $\mathcal{A}$ and $\mathcal{B}$ use the FS setup, and $\mathcal{C}$ and $3\mathcal{D}$  are traditional RW shock simulations. $\Delta x$ and $\Delta t$ are the number of cells per ion skin depth, and the number of cyclotron times per simulation timestep, respectively.}
\label{table:sim_parameters}
\end{table*}

We implement our setup with the relativistic hybrid code \texttt{dHybridR} which uses a particle-in-cell approach with fully kinetic ions and a charge-neutralizing electron fluid \citep{gargate+07, haggerty+19a}. 
Typical shock simulations using a reflecting wall (hereafter: RW) produce a shock by reflecting a super-sonic flow off a stationary wall and letting it interact with the incoming flow; 
sometimes this is implemented with a conducting piston moving through a fluid at rest \citep[e.g.,][]{gupta+24b}, but this is only equivalent to a moving RW. 
In our setup the left boundary of the box adopts a semi-permeable, semi-reflecting boundary condition, which we dub the \textit{faux shock}, FS.
The FS acts as a open boundary for thermal particles and as a shock for a separate population of suprathermal particles (hereafter, the CRs), i.e., allowing a given fraction of the CRs to leave the box and reflecting the rest as if they have been scattered in the downstream. 
Assuming that CRs are isotropic in the downstream, the probability that a particle crossing the left hand boundary returns to the simulation domain reads  \citep{peacock81}:
\begin{equation}
    P_{\rm return}=\left(\frac{1-u_2/v_{\rm cr}}{1+u_2/v_{\rm cr}} \right)^2 ,\label{shock_reflect_condition}
\end{equation}
where $v_{\rm cr}$ and $u_{2}$ are the CR and downstream fluid speeds, respectively, as measured in the shock frame.
To first order in $u_2/v_{\rm cr}$ eq. (\ref{shock_reflect_condition}) gives $P_{\rm return}  \approx  1 - \frac{4 u_{2}}{v_{\rm cr}}$, in agreement with the result of \citet{bell78a}. 
This can similarly be written to highlight its dependence on the shock compression ratio, $r$, by substituting $u_2 = u_{\rm sh}/r$. 

Because the simulation takes place in the shock frame, CRs ---on average--- lose energy when scattering in the downstream due to tail-on collisions, but gain energy in their upstream scattering via head-on collisions so that there is a net increase of energy on every cycle. Following \cite{bell78a}, the average energy gain per cycle is: 
\begin{equation}
    \left< \frac{\Delta E}{E} \right> \approx \frac{4}{3} \frac{u_1 - u_2}{c} = \frac{4}{3} \frac{u_1}{c}\left( 1 - \frac{1}{r} \right)
    ,\label{energy_gain_equation}
\end{equation}
giving rise to a differential particle momentum spectrum
\begin{equation}
    \frac{dN(p)}{dp} = 4 \pi p^2f(p) \propto p^{-q_{\rm p}}
\end{equation} 
with spectral index
\begin{equation}
    q_{\rm p} = \frac{3r}{r-1}
\end{equation} 
where for $r=4$, $q_{\rm p}=4$, and $dN/dp \propto p^{-4}$.
Importantly, $q_{\rm p}$ depends on the compression ratio, and this quantity is fully customizable within our simulation setup. 

The right boundary of the simulation domain is open for energetic particles to escape to upstream infinity, and both transverse dimensions are periodic. 
The thermal background plasma is injected continuously from the right boundary of the box, and it moves leftward toward the FS with speed $u_{\rm sh}$. 
CRs are constantly injected as a second species from the left hand boundary with a starting momentum, $p_{\rm iso}$, which is isotropic in the downstream frame, and moving in bulk with the downstream flow at velocity $u_2$.
Within the isotropic distribution, some particles will have insufficient $p_x$ to overcome $u_2$, so some additional attention is necessary to inject the desired initial CR number density.


We initialize the box with an ordered magnetic field $\mathbf{B}_0$; 
once CRs begin entering the box, they drive a current that modifies and amplifies the initial fields, producing the fluctuations that cause them to scatter, return to the shock, and eventually trigger self-sustained DSA. 

This new simulation setup gives up the ability to capture injection self-consistently but has several advantages: 
1) we inject an initial population of CRs at a fixed momentum and number density, calibrated on global shock simulations \citep[e.g.][]{orusa+23}, effectively jump-starting the Bell instability;
2) the simulation is in the shock frame and all the computational power is focused on the physics of the upstream--- this allows us to simulate precursors on scales previously inaccessible;
3) injecting CRs as a separate species allows us to drastically improve their statistics--- quantitatively, since the fraction of particles injected into DSA never exceeds $\sim 1\%$ \citep{caprioli+15}, using the same number of thermal and CR macroparticles per cell effectively increases the CR phase-space resolution by two orders of magnitude with only twice as many particles;
4) this framework can also be applied to simulate the re-acceleration of pre-existing CRs, and to model shock configurations in 2D that otherwise need 3D physics to accurately capture ion injection \citep{jones+98, orusa+23}.

To test the accuracy of the FS setup, we compare our results to a standard RW shock simulation with the same parameters.
The benchmark FS simulation, referred to as Run $\mathcal{A}$ in Table \ref{table:sim_parameters}, and the RW simulation, referred to as Run $\mathcal{C}$, feature a 2D box with sizes $10\,000$ by $300$ $d_{\rm i}$ and $20\,000$ by $300$ $d_{\rm i}$ in the $x$ and $y$ directions respectively, where $d_{\rm i}\equiv c/\omega_{\rm p}$ is the ion skin depth and $\omega_{\rm p} = \sqrt{4\pi n_{\rm g} e^2/m}$ is the ion plasma frequency, with $c$ the speed of light, $m$ the ion mass, $e$ the ion charge, and $n_{\rm g}$ the ion number density.
The RW simulation is given a larger length in $x$ to allow for the propagation of the shock.
The initial magnetic field $\mathbf{B}_0=B_0 \mathbf x$ is oriented parallel to the shock normal $\mathbf{n}$ (forming an angle $\thbn = 0^{\circ}$) and anti-parallel with the thermal plasma velocity. 
Run $\mathcal{A}$ has an Alfv\'enic Mach number $M_{\rm A} = u_{\rm sh}/v_{\rm A} =26.67$ as seen from the shock frame, with $v_{\rm A} \equiv B_0/\sqrt{4\pi m n_{\rm g}}$ the Alfv\'en speed. 
Run $\mathcal{C}$, the corresponding RW simulation, is in the downstream frame and its upstream fluid is initialized with speed $M_{\rm A} = 20$, such that both simulations have the same effective $M_{\rm A}$ in the shock frame assuming that the shock compression ratio remains constant.
Time is measured in inverse cyclotron times, $\omega_{\rm c}^{-1} \equiv mc/eB_0$, speed in units in $v_{\rm A}$, and energy in units of $m v_{\rm A}^2$. 
In Run $\mathcal{A}$, CRs are injected with number density $n_{\rm cr} = 0.01 n_{\rm g}$ and with a momentum distribution isotropic in the downstream frame, and each CR has momentum $p_{\rm iso} = 100 mv_{\rm A} = 0.1c$. 
Both simulations are run for about $450 \, \omega_{c}^{-1}$, which is sufficient to ensure that CRs develop a non-thermal spectrum and the magnetic field amplification driven by the Bell instability has saturated.

\section{Comparing Simulations with FS and RW}\label{sec:results}

The FS setup is intrinsically different from a traditional RW simulation setup, meaning their respective outputs are not expected to be identical. 
The following are important considerations to keep in mind in order to meaningfully compare results obtained with the FS setup with those from RW simulations. 

\textit{Frame shifting.} FS simulations are performed in the shock frame, whereas the presented RW simulations are performed in the downstream frame. Physical quantities therefore need to be boosted for comparison. 
In the limit where the fluid speed is $u \ll c$, a simple Galilean transformation can be applied to the particle spectra and electromagnetic fields change at the order of $\mathcal{O}( u/c)$.

\textit{Time offset.} The timesteps in a FS simulation will not correspond one-to-one with the timesteps in a RW simulation. 
This is because CRs are injected at the FS with a given $p_{\rm iso}$, which anticipates the $p_{\rm inj}$ found in RW simulations; 
thus, the FS will systematically be ``ahead" of the RW simulations for the same timestep. 
This creates a time offset $\Delta t = 150 \, \omega_c^{-1}$ between the parallel-shock simulations which we have checked persists as a linear shift for the duration of our runs.
Therefore, we compare the FS and RW output at the same evolutionary stage where the FS is shown at $\Delta t$ earlier than the RW.

\textit{Comparing just upstream quantities.} In RW simulations there is a downstream (left) and an upstream (right) separated by the shock which physically propagates from left to right in time. 
The FS has only an upstream, and therefore we will compare spectra in the respective upstreams, placing the shock/FS against the leftmost edge.

\textit{Separate thermal and non-thermal species.} In traditional RW simulations, it is not possible to separate the CRs from the thermal plasma without making cuts in energy. 
However, the FS setup is designed to treat the thermal plasma and the CRs as two separate species and can easily isolate the CR component. 
Therefore, plots of the FS results will show just the CR component, whereas plots of the RW simulation will show a mixture of CRs and thermal particles. 
When presenting results, we will draw emphasis only to the comparison between the non-thermal components of each simulation. 

\textit{CR injection efficiency.} We prescribe our CR normalization for the FS setup based on previous hybrid runs, and validate our choice by comparing the normalization of the FS and RW's non-thermal particles. 

\textit{Early CR transient.} Due to the nature of injecting our first population of CRs into a ``pristine" upstream, these CRs form a ``transient" in the sense that they mostly propagate unscathed into the far upstream and escape through the open boundary on the right edge. 
These CRs provide the initial current which seeds the Bell instability and causes magnetic turbulence which leads to DSA. 
This transient is likely unphysical in that it carries an overdensity of CRs at high energies, and thus produces an unphysically high current. 
Because of this, we wait for the transient to propagate all the way through the box and exit through the open right-side boundary. 
After this initial transient phase, the system relaxes and all currents and fields converge to the values self-consistently produced by DSA. 

We will take all of these effects into account when discussing the FS vs RW simulations. 
A list of all runs discussed in this paper can be found in Table \ref{table:sim_parameters}.

\subsection{Parallel Shocks}
\label{ssec:parallel}

\begin{figure*}[h!]
    \centering
    \subfigure[$p_{tot}$]{
        \includegraphics[trim = 15 30 32 5, clip=true, width=0.45\textwidth]{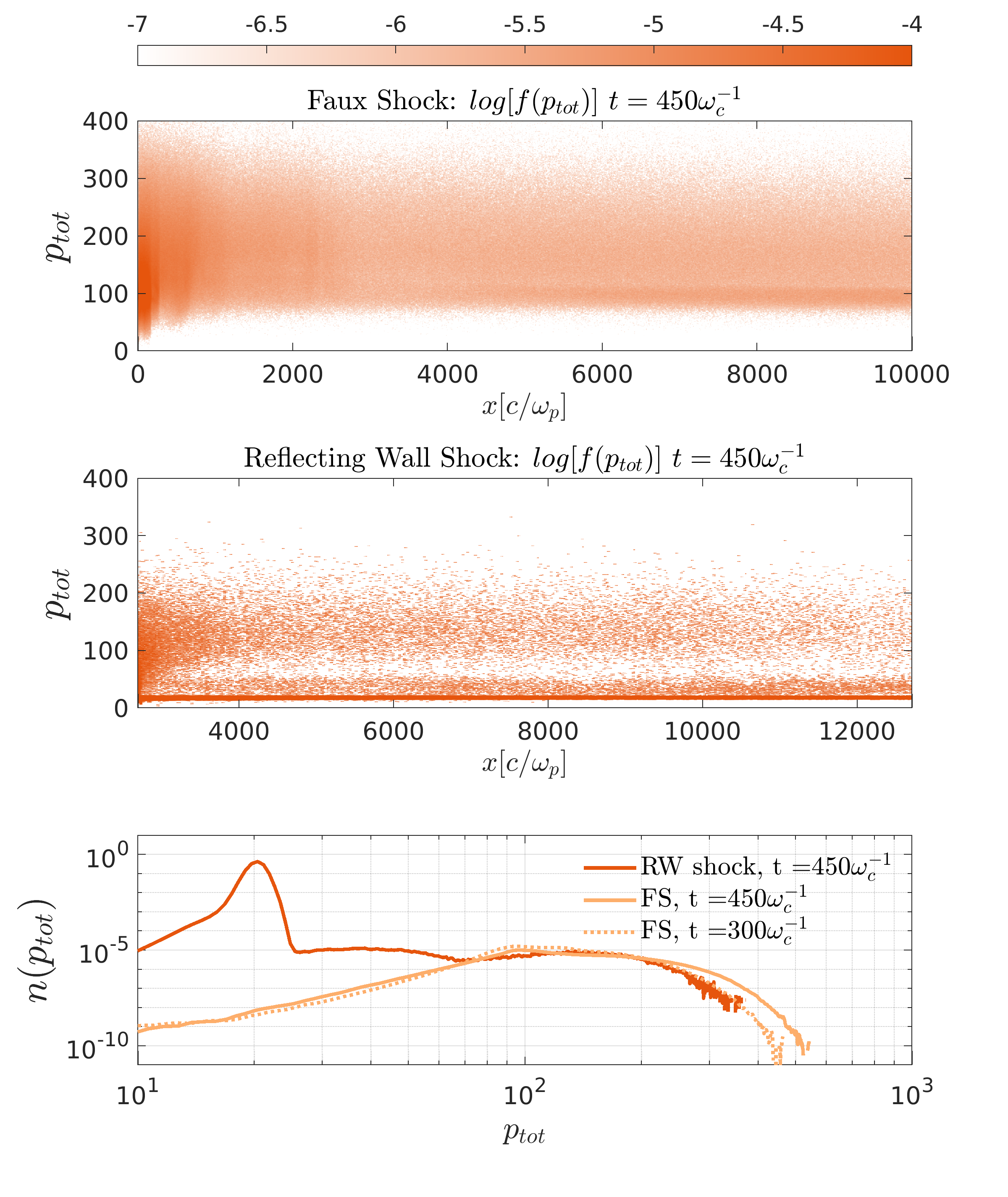}
    }
    \hspace{-0.3cm}
    \subfigure[$p_x$]{
        \includegraphics[trim = 15 30 32 5, clip=true, width=0.45\textwidth]{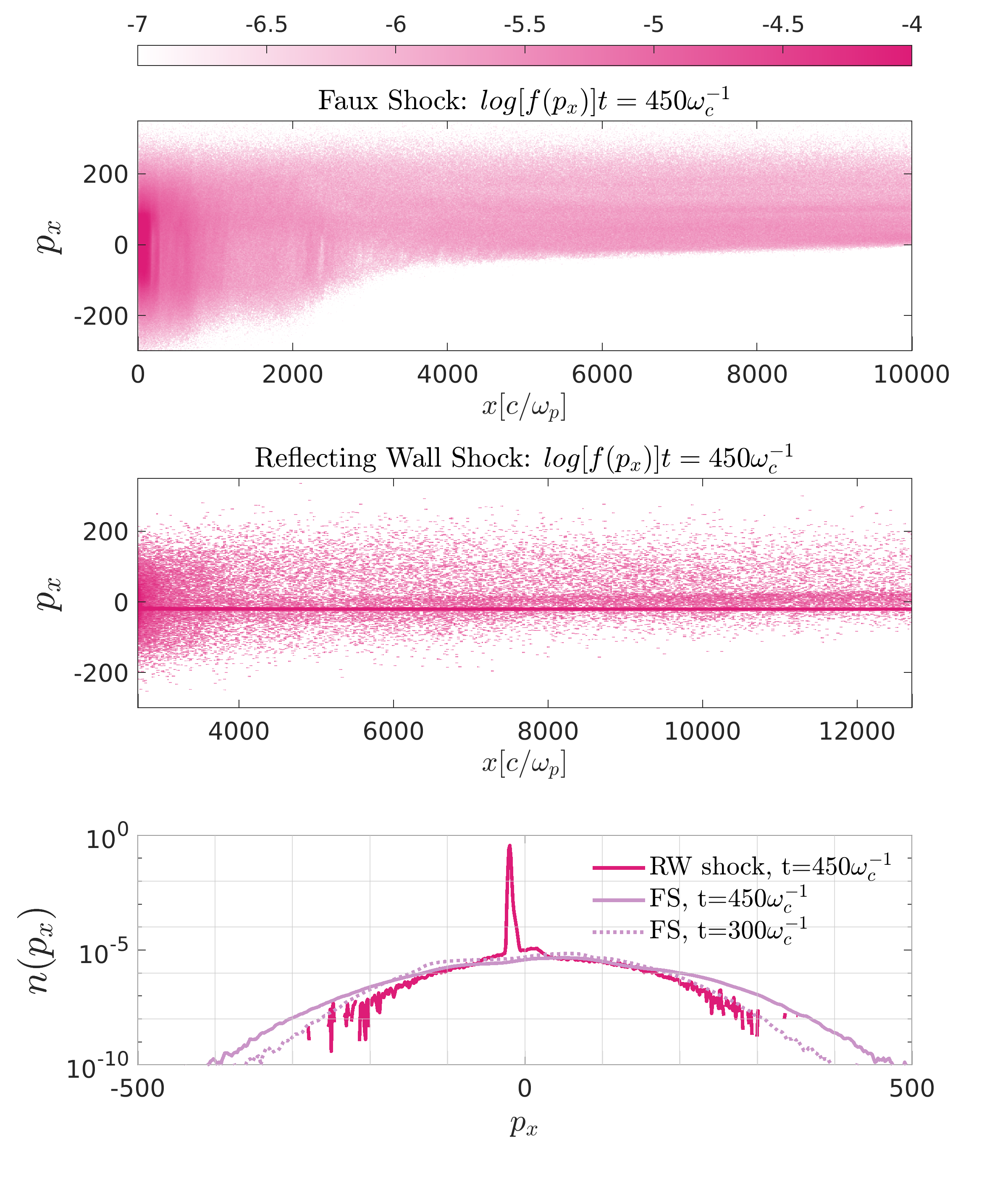}
    }
    \subfigure[$p_y$]{
        \includegraphics[trim = 15 30 32 5, clip=true, width=0.45\textwidth]{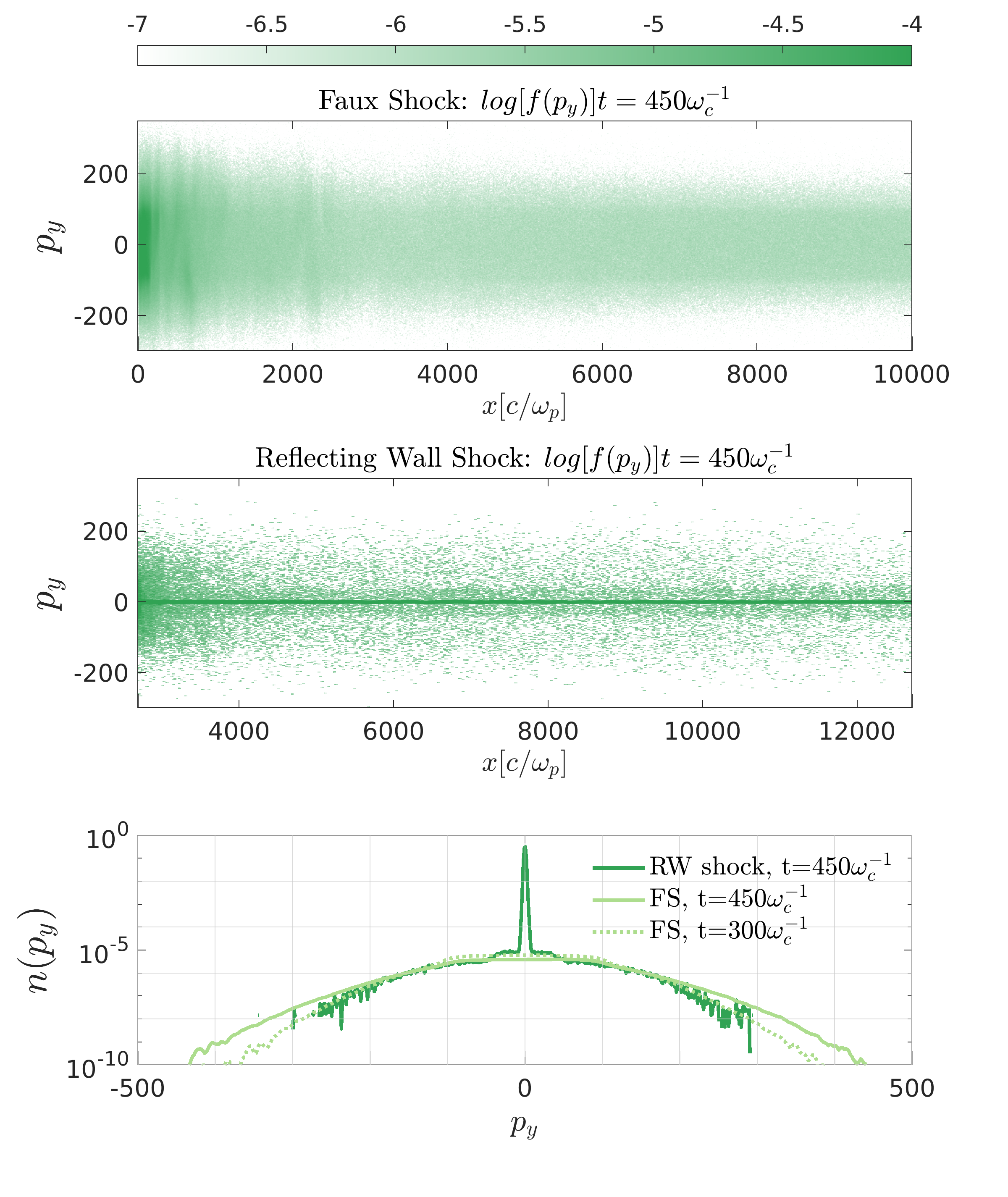}
    }
    \hspace{-0.3cm}
    \subfigure[$p_z$]{
        \includegraphics[trim = 15 30 32 5, clip=true, width=0.45\textwidth]{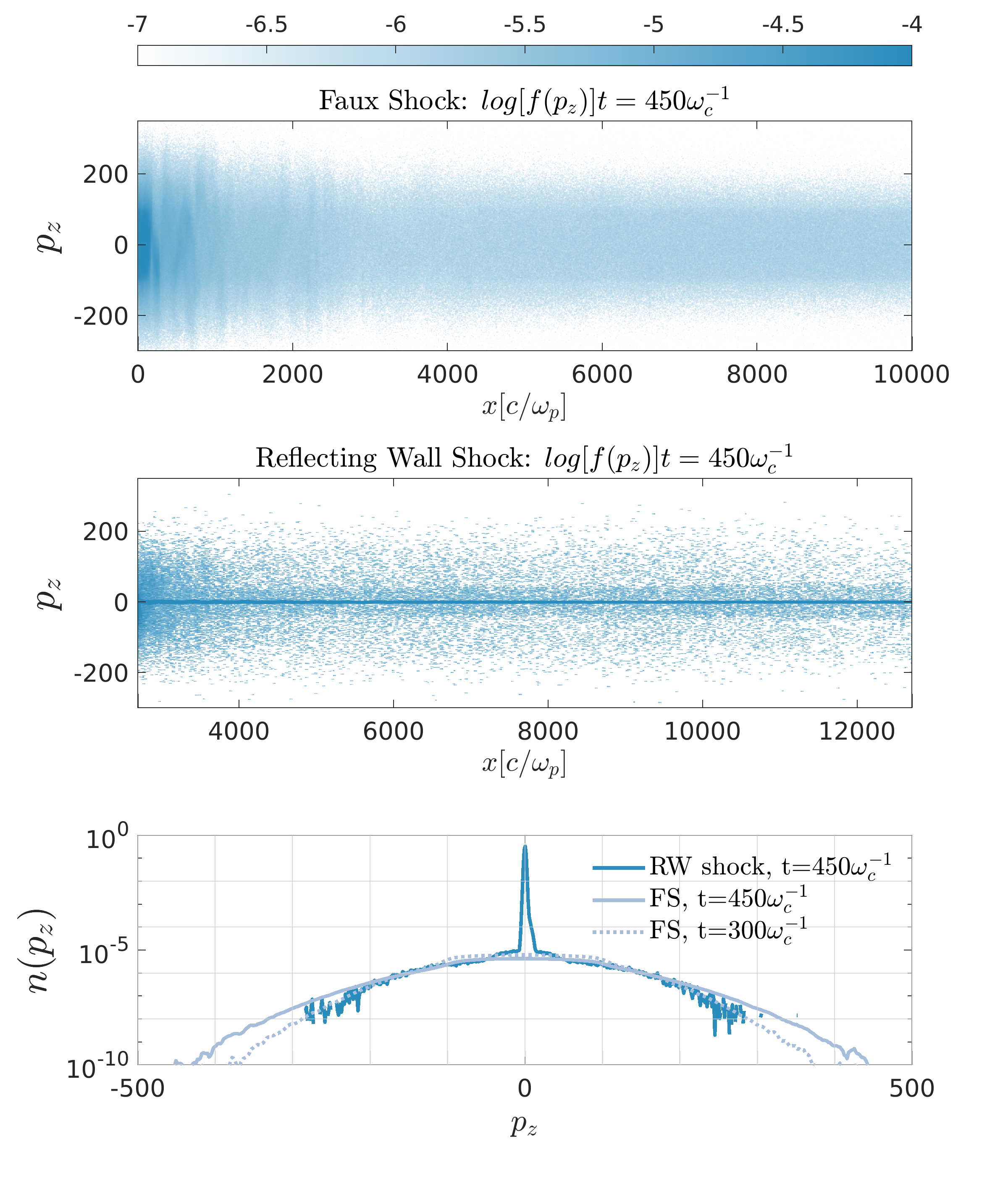}
    }
    \caption{Momentum phase space in units of $mv_{\rm A}$ of the benchmark FS simulation, Run A (top panels), compared to a RW shock simulation, Run C (middle panels) at $t=450 \, \omega_{\rm c}^{-1}$. The bottom panels show a comparison of the momentum spectra between both runs at the same simulation timestep (solid lines) and the FS at the same relative time in the shock's evolution (dotted lines).}
    \label{fig:momentum_RunA}
\end{figure*}

We start by considering our benchmark FS Run $\mathcal{A}$ and its RW counterpart, Run $\mathcal{C}$, each with upstream velocity $v_1 = 20 v_{\rm A}$ as seen from the downstream frame, with effective $M_{\rm A} = 20$ in the perspective on the downstream.
Figure \ref{fig:momentum_RunA} shows the momentum phase space distributions of Run $\mathcal{A}$ and $\mathcal{C}$ (top and middle panels) for the total momentum (subfigure a) and for the individual components, $p_x$, $p_y$, and $p_z$ respectively (subfigures b-d).
Bottom panels show a comparison of the momentum spectra in each case, and the FS is shown at the same simulation timestep and with the $\Delta t = 150 \, \omega_c^{-1}$ offset.
At $t=450\,\omega_{\rm c}^{-1}$, the shock in the RW simulation is at $x \approx 2700 \, d_{\rm i}$, and we plot just the upstream region at larger $x$.
The FS panels of Figure \ref{fig:momentum_RunA} show the isolated CR component of the phase space distributions whereas the RW shock panels show CRs and the thermal plasma together.

\subsubsection{Momentum Spectra}
The phase space distributions of Run $\mathcal{A}$ and Run $\mathcal{C}$ show similar CR populations, with Run $\mathcal{C}$ (the FS) resolving finer features due to its higher CR resolution.
We verify that our prescribed normalization of the CRs in Run $\mathcal{A}$ ($n_{\rm cr}/n_{\rm g} = 1\%$) matches the injection in Run $\mathcal{C}$ by comparing the spectra in the bottom panel and seeing that they indeed reach the same $n(p)$ in the non-thermal tail. 
Spectra are compared both at the same simulation timestep ($t=450 \, \omega_{\rm c}^{-1}$) (solid lines), and at the same time relative to the FS evolution ($t=300 \,\omega_{\rm c}^{-1}$) (dotted lines), which occurs at a constant time offset of $\Delta t = 150 \, \omega_{\rm c}^{-1}$. 
The offset time comparison allows us to see the FS and RW shocks at the same evolutionary stage, allowing for direct comparison of their spectra; 
the FS spectra at a given time forecasts the future RW spectra, effectively reducing the computational cost of the simulation.

Subfigure \ref{fig:momentum_RunA}(b) demonstrates that DSA is taking place in both setups.
In the FS case, CRs are injected at  $t=0 \, \omega_{\rm c}^{-1}$ from the left boundary with momentum in the $+x$ direction and the upstream contains no magnetic field turbulence beyond the initial numerical noise.
As CRs drive a current into the upstream, magnetic fluctuations develop because of the Bell instability, scattering CRs back towards the shock with $-p_{x}$.
At $t = 450 \, \omega_{\rm c}^{-1}$, there is one population of particles diffusing within $x \sim 2500 \, d_{\rm i}$ from the shock, with a distribution essentially isotropic (almost symmetric in $p_x$ ), and one population freely streaming away from the shock with positive $p_x$ propagating towards upstream infinity.
This matches findings from the RW shock simulations (see the middle panel).

Subfigures (b) and (c) show the momentum phase space distributions in $p_y$ and $p_z$, which are symmetric around $0$, as we would expect from an initial magnetic field in the $x-$direction; they also agree well between the RW and FS cases.


\subsubsection{Magnetic Fields}
\begin{figure*}
    \centering
    \subfigure[$B_{tot}$]{
        \includegraphics[trim = 27 40 35 13, clip=true, width=0.45\textwidth]{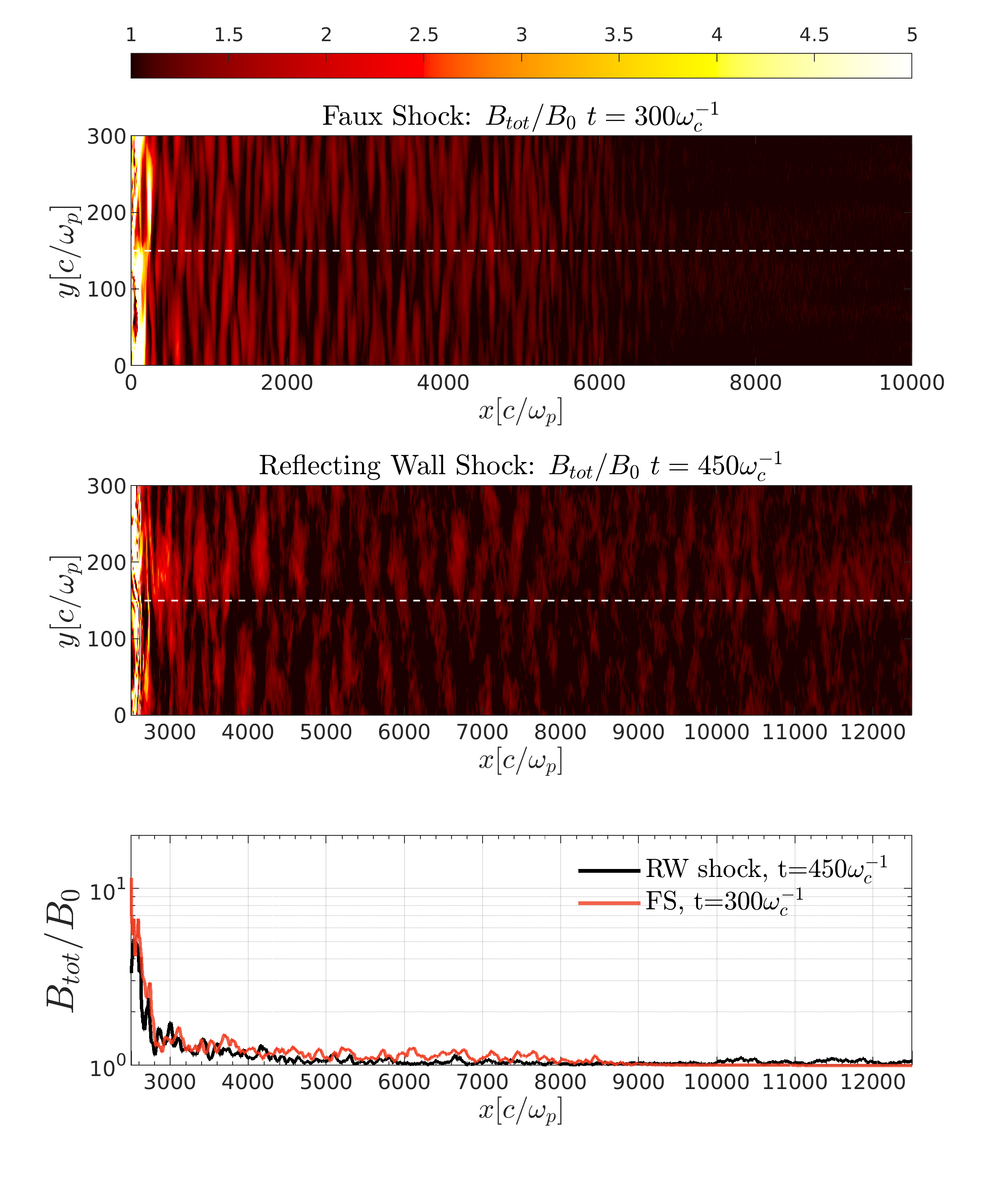}
    }
    \hspace{-0.3cm}
    \subfigure[$B_x$, $B_y$, $B_z$]{
        \includegraphics[trim = 20 15 40 40, clip=true, width=0.45\textwidth]{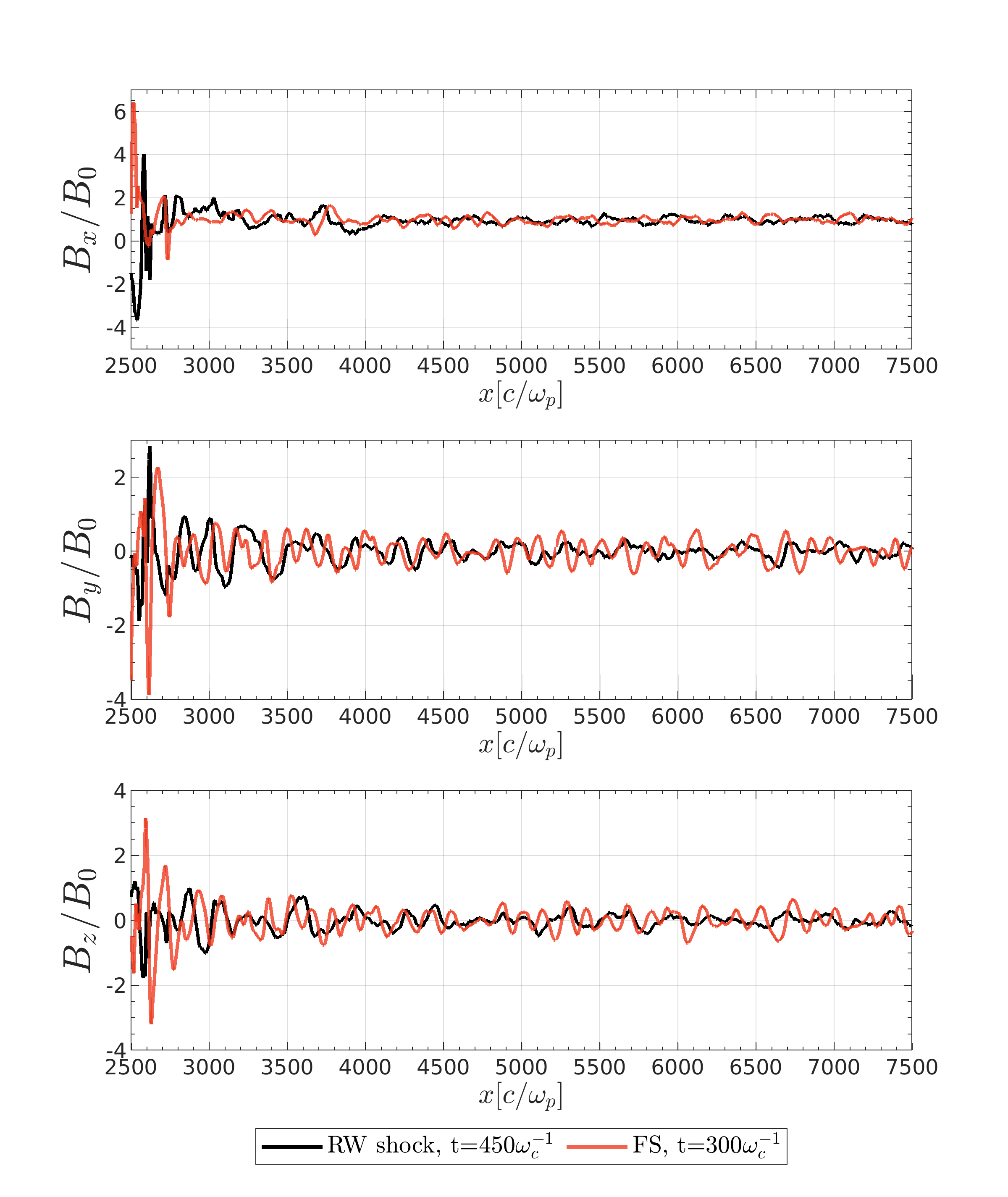}
    }
    \caption{(a) Comparison of the total amplified magnetic field strength of the benchmark FS simulation (top panel) versus the RW shock simulation (middle panel) at the same simulation timestep, together with lineouts (taken from $y=25 \, d_{\rm i}$, dashed white lines) of the magnetic field strength (bottom panel). (b) Lineout comparisons of each component of the magnetic field ($B_x$, $B_y$, and $B_z$) for the benchmark FS simulation (red) versus the RW shock simulation (black).}
    \label{fig:B_comps_RunA}
\end{figure*}

Figure \ref{fig:B_comps_RunA}(a) shows the total magnetic field in units of the initial field strength at $t = 300$ and $t = 450 \, \omega_{\rm c}^{-1}$ for the FS and RW shock respectively, over $10\,000 d_{\rm i}$ upstream of each shock. 
The top two panels show the 2D maps, while the bottom panel shows the fields averaged over the $y$ dimension; 
the FS curve (red) has been shifted right by the shock displacement, $2700$ $d_{\rm i}$, so that the upstream regions of both simulations overlap.
Both the FS and RW show similar features: $\delta B/B_0\sim 5-10$ close to the shock, and then its value tapers down to $\sim 1.5$ in the extended shock precursor. 
Additionally, the wavelength and transverse sizes of the semi-coherent wave structures driven by the Bell instability are roughly the same by eye, $\sim 100 \, d_{\rm i}$. 

However, we notice that Run $\mathcal{C}$ has small-scale fluctuations farther upstream than Run $\mathcal{A}$; 
this is a consequence of Run $\mathcal{C}$'s larger initial box length-- a necessary condition to offset the shock propagation that occurs in all RW shock setups.
The Bell instability begins growing from scratch at the right edge of the FS's box ($10\,000 d_{\rm i}$), and equally from the right edge of the RW shock's box ($20\,000 d_{\rm i}$), meaning some fluctuations are already visible throughout the truncated box shown.


Subfigure \ref{fig:B_comps_RunA}(b) shows horizontal slices (lineouts) of the separate components of the magnetic field, $B_x$, $B_y$, and $B_z$, compared to the initial field strength, $B_0$, at $t = 300$ and $t=450 \, \omega_{\rm c}^{-1}$ for the FS and RW shock respectively. 
Overall, these profiles show that the FS framework can capture the development of the Bell instability in the shock precursor in close agreement with the RW setup.

\begin{figure}[h!]
    \centering
    \subfigure[FS]{
        \includegraphics[trim = 14 1 18 14, clip=true, width=0.45\textwidth]{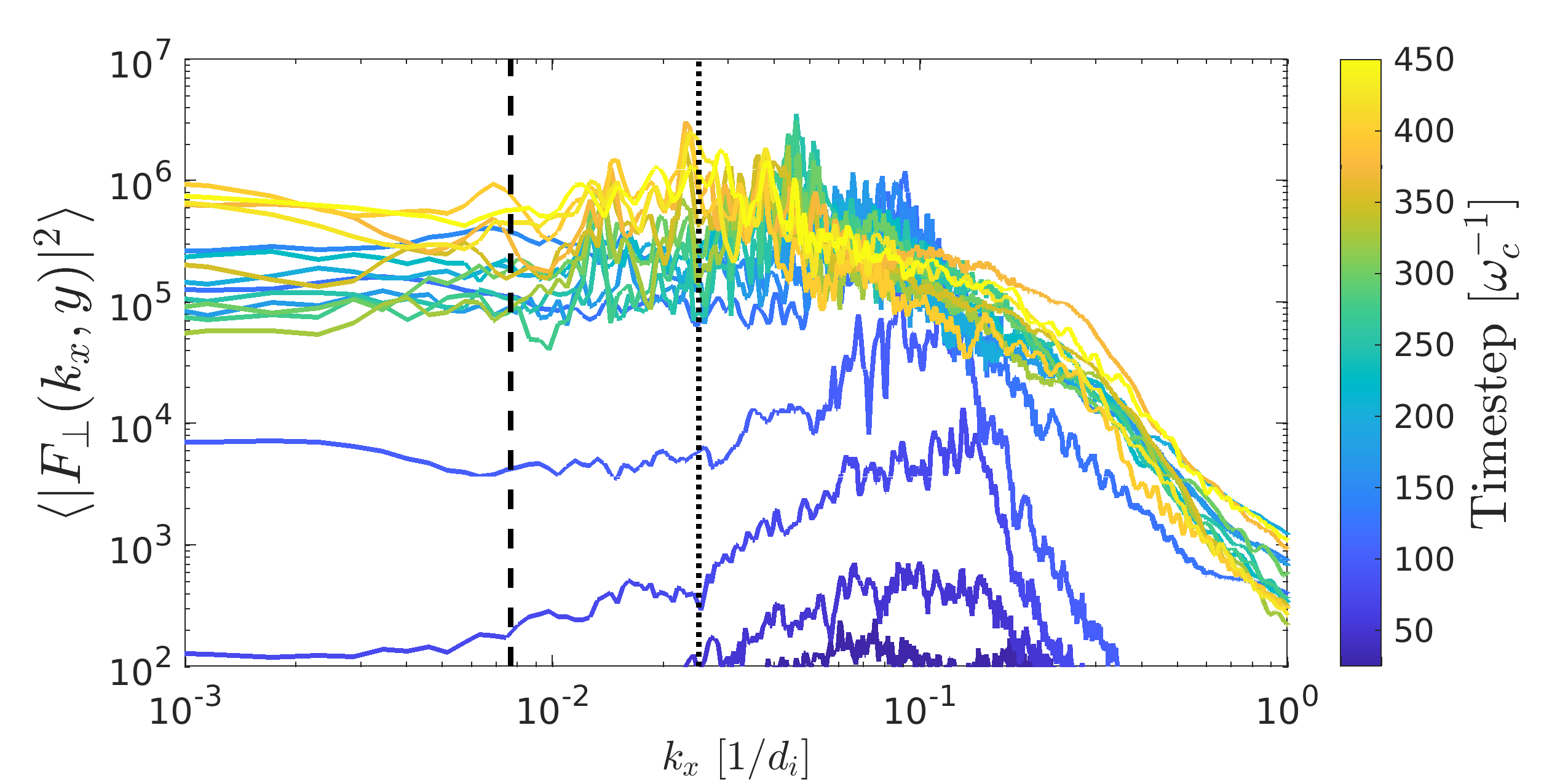}
    }
    \vspace{-0.3cm}
    \subfigure[RW]{
        \includegraphics[trim = 14 1 18 14, clip=true, width=0.45\textwidth]{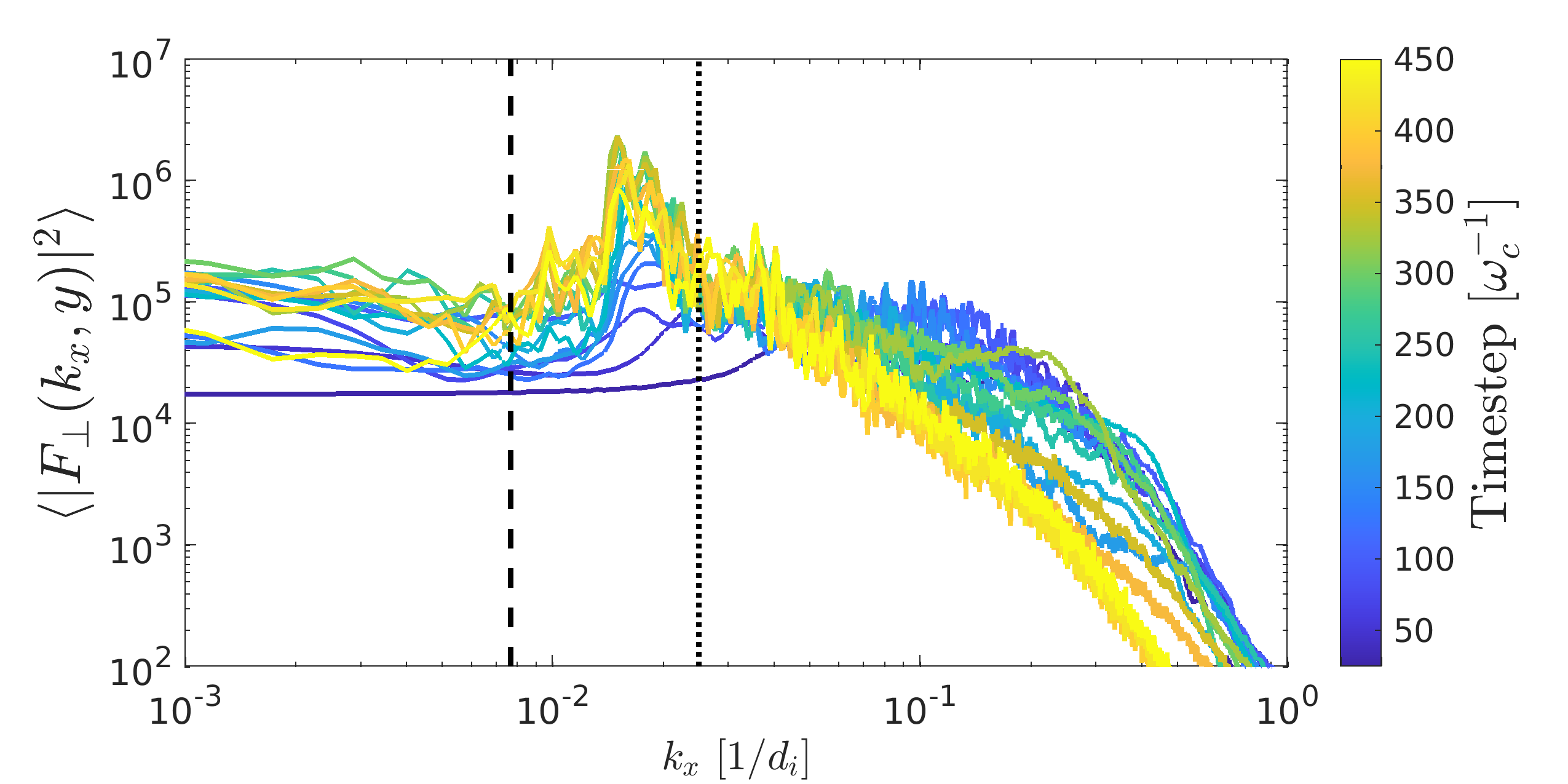}
    }

    \caption{Perpendicular magnetic power spectrum averaged over the transverse direction, calculated as $\left<|\mathcal{F}_{\perp}(k_x,y)|^2 \right> = |\mathcal{F}[B_y(x,y)]|^2 + |\mathcal{F}[B_z(x,y)]|^2$ for the FS (a) and RW (b) over time. The black, dashed line marks the mode resonant with CRs with $E_{\rm max} = 130 E_{sh}$. The black, dotted line marks $k_{\rm max}$ predicted from the Bell instability for a CR current $J_{\rm CR}\sim 0.05 \, n_{\rm g} v_{\rm A}$.}  
    \label{fig:FT}
\end{figure}

Decomposing the Fourier modes of the magnetic field provides further insight into the nature of the waves.
Figure \ref{fig:FT} shows the transversely-averaged perpendicular magnetic power spectrum as a function of wavenumber in units of $1/d_{\rm i}$ for both the FS (a) and RW shock (b) setups over time (color-coded). 
The Bell instability predicts a fastest-growing mode with wavenumber $k_{\rm max} = \frac{4 \pi}{c} \frac{J_{\rm CR}}{B_0}$, where $J_{\rm CR}$ is the current carried by CRs \citep{bell04}.

During the early linear stage in the FS simulation, the initial transient produces a relatively large CR current ($J_{\rm CR} \sim 0.2 n_{\rm g} v_{\rm A}$), corresponding to $k_{\rm max} \sim 0.1 \, d_i^{-1}$.
This wavenumber aligns with the prominent early-time peaks observed in the FS power spectrum (subfigure a, blue curves).
By contrast, these peaks are not seen in the RW simulation (subfigure b), consistent with its lack of a comparable transient.

As DSA progresses in the FS, CRs scatter and become trapped, lowering the upstream current.  
Previous estimates suggest that the Bell instability dominates over the resonant instability near $M_{\rm A} \approx 30$ \citep{caprioli+14b}.
For $M_{\rm A} = 26.7$ (in the shock frame) as is the case for Runs $\mathcal{A}$ and $\mathcal{C}$, growth is expected between the Bell and resonant scales. 
At late times the current that permeates most of the upstream ($x > 1000 \, d_{\rm i}$) is roughly constant at the level of $J_{\rm CR} \sim 0.05 n_{\rm g} v_{\rm A}$, which returns $k_{\rm max} \sim 2.5 \times 10^{-2} d_i^{-1}$ (black dotted line).
A more slowly-growing mode also appears near the resonance scale of the highest energy CRs ($E_{\rm max} \sim 130 \, E_{\rm sh}$), corresponding to $k_x \sim 1/r_{\rm g}\sim 7.7 \times 10^{-3} \, d_{\rm i}^{-1}$ (black dashed line), where $r_{\rm g}$ is the gyroscale of CRs at $E_{\rm max}$.
Accounting for the time offset of $150 \, \omega_{\rm c}^{-1}$, both setups show comparable power at the largest scales, $k_x \lesssim 0.1 \, d_{i}^{-1}$.
At smaller scales, differences arise, particularly in modes powered by particles with the lowest energy particles. 
These discrepancies are consistent with the FS setup tracking only injected CRs, all initialized at $p_{\rm iso} = 100 \, mv_{\rm A}$.

\subsection{Quasi-perpendicular Shocks}
\label{ssec:oblique}
\citet{orusa+23} found that particle injection at highly oblique shocks ($\theta_{\mathrm{Bn}} = 80^\circ$) is possible in 3D simulations, whereas little to no injection occurs in 2D simulations with otherwise identical parameters,  highlighting the importance of magnetic field line wandering along the third dimension \citep{jones+98}.
3D simulations increase computational cost sharply, thus imposing limits on the size of the box and the duration of the simulations. 
The FS setup can jumpstart ion injection, meaning that injection can be set as an initial condition in 2D boxes, drastically reducing the computational cost and providing access to regimes that are physically interesting but have been practically impossible to simulate with traditional methods.

\begin{figure*}[h!]
    \centering
    \subfigure[$p_{tot}$]{
        \includegraphics[trim = 15 30 32 5, clip=true, width=0.45\textwidth]{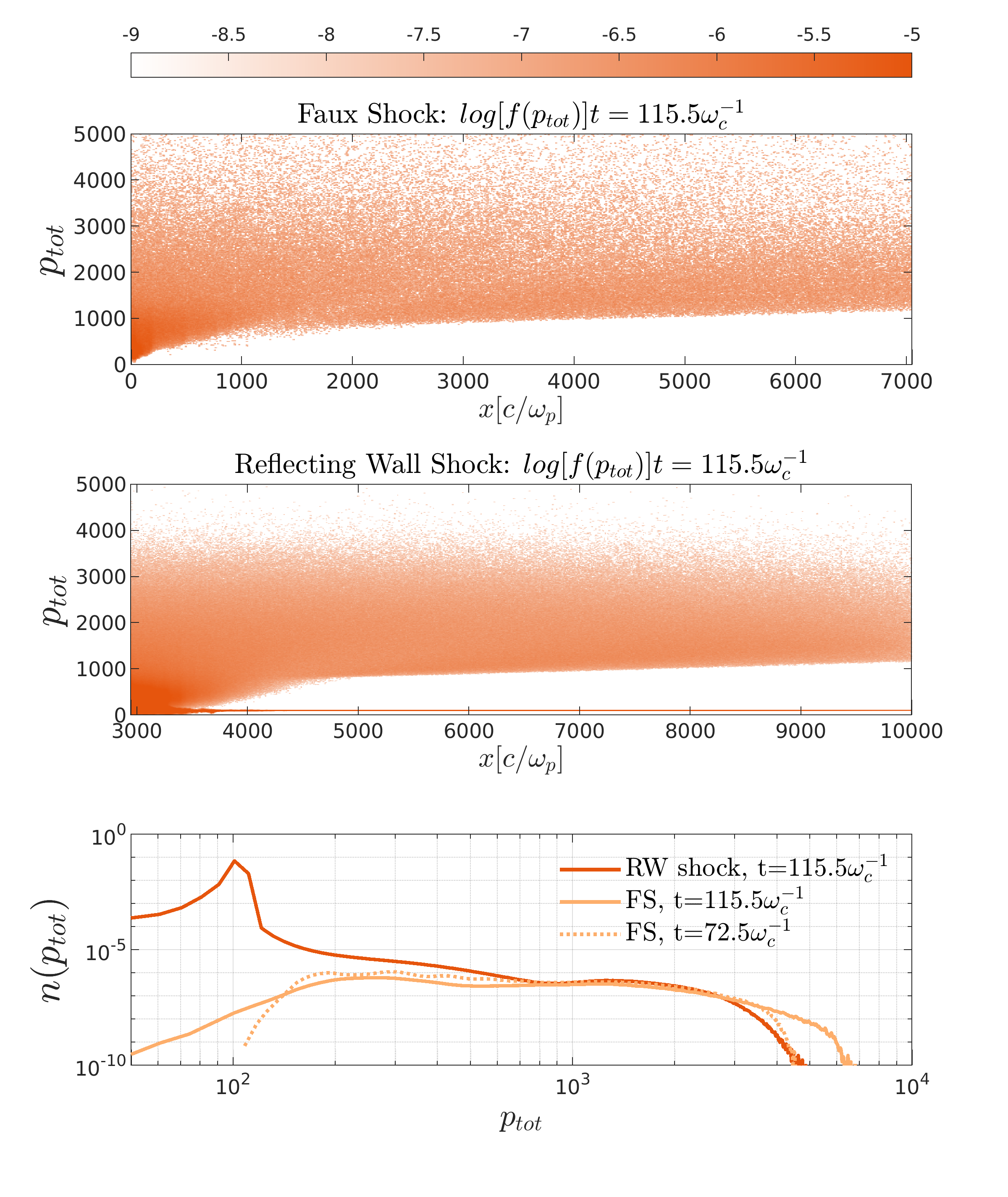}
    }
    \hspace{-0.3cm}
    \subfigure[$p_x$]{
        \includegraphics[trim = 15 30 32 5, clip=true, width=0.45\textwidth]{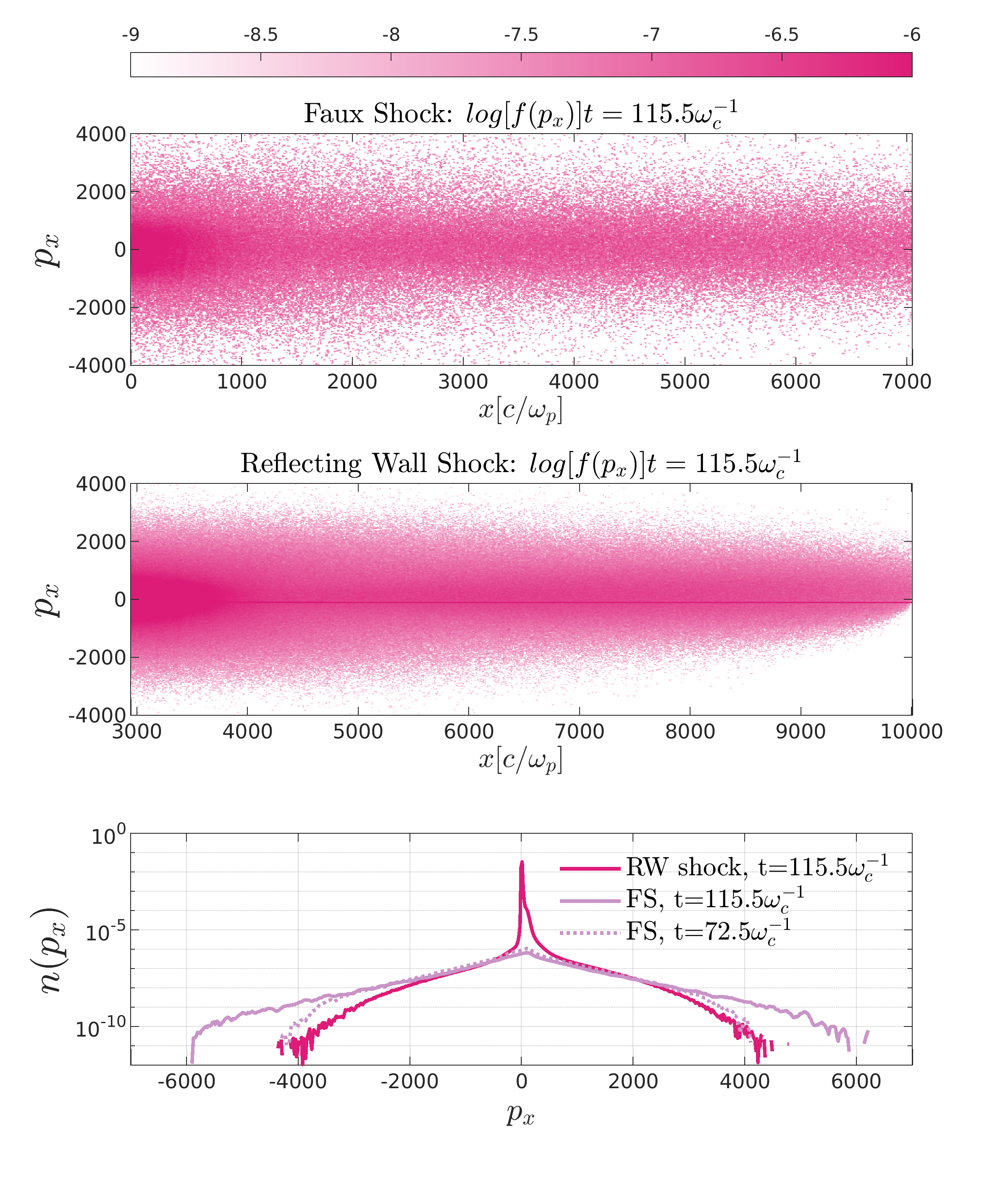}
    }
    \subfigure[$p_y$]{
        \includegraphics[trim = 15 30 32 5, clip=true, width=0.45\textwidth]{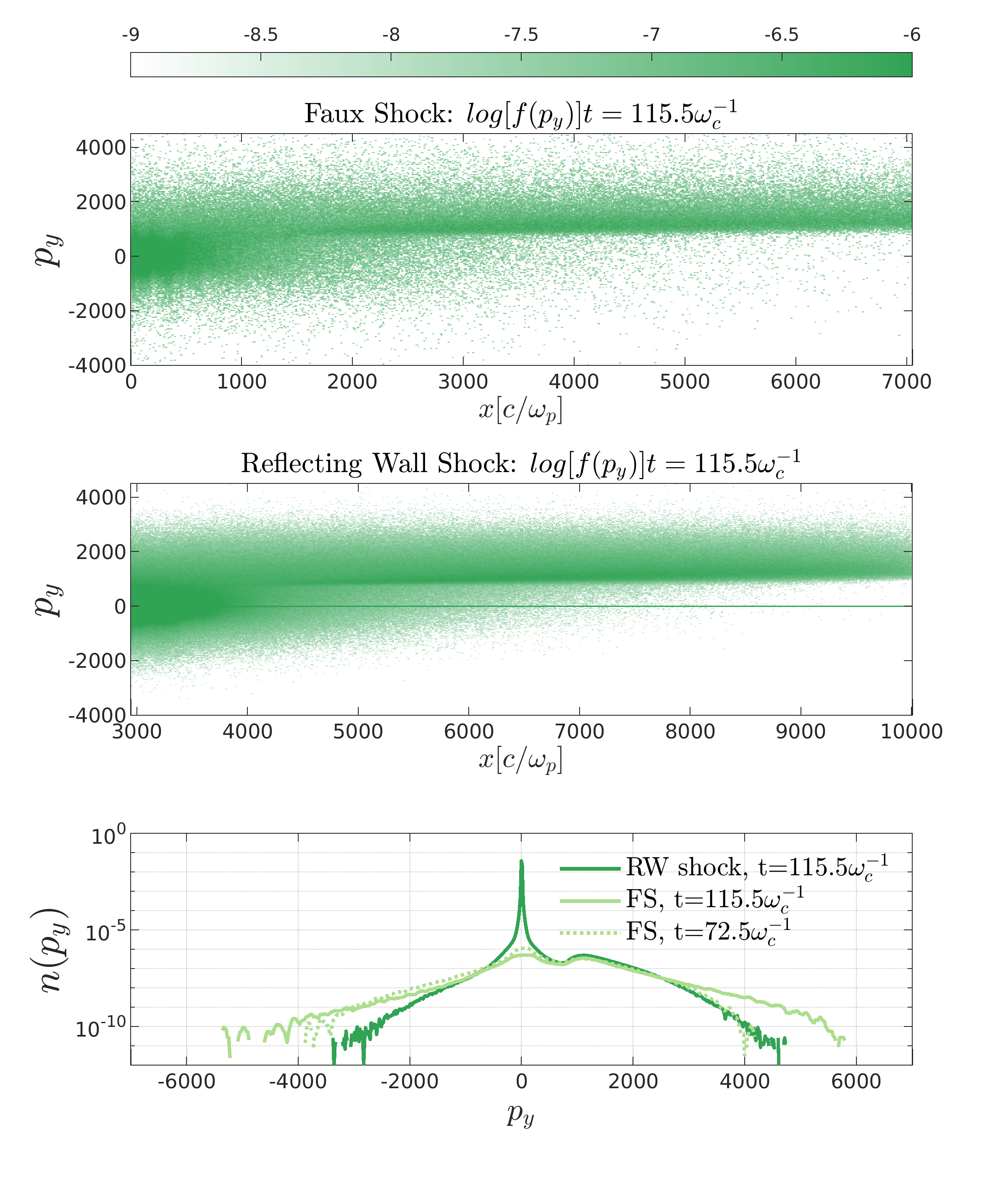}
    }
    \hspace{-0.3cm}
    \subfigure[$p_z$]{
        \includegraphics[trim = 15 30 32 5, clip=true, width=0.45\textwidth]{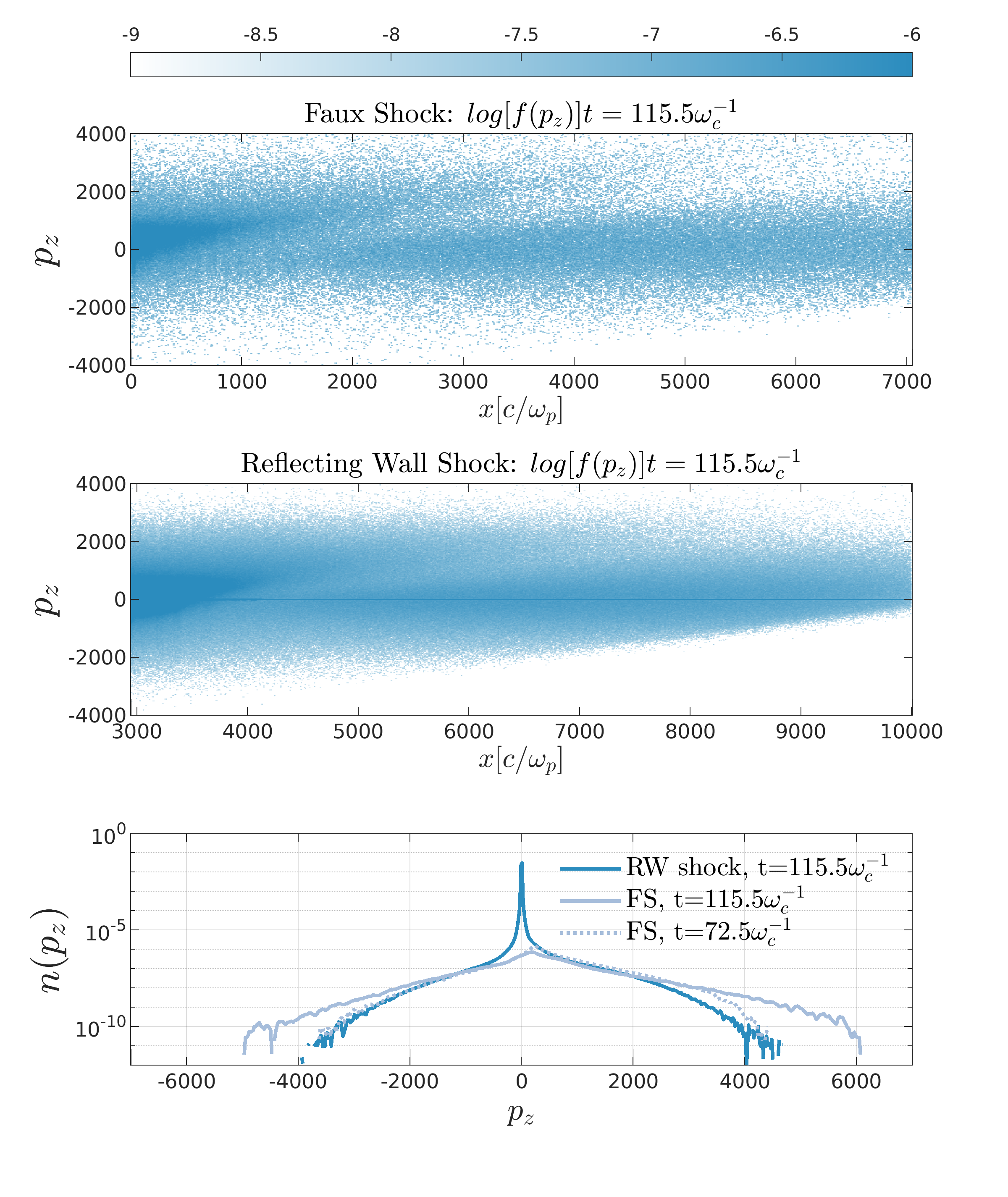}
    }
    \caption{Momentum phase space of the $80$ degree FS simulation (Run B) compared to a 3D RW shock simulation (Run $3\mathcal{D}$) at $t=115.5 \,\omega_{\rm c}^{-1}$. (a) The total momentum phase space of the FS (top panel), the $3\mathcal{D}$ RW shock (middle panel), and the spectrum of each compared at the same simulation timestep (solid line) and relative time in the shock's evolution (dotted line) (bottom panel). (b-d) The same as (a) but for $p_x$, $p_y$, and $p_z$.}
    \label{fig:momentum_RunB}
\end{figure*}

The canonical $p^{-4}$ spectrum associated with DSA at strong shocks is not guaranteed at quasi-perpendicular shocks \citep[e.g.,][]{bell+11}, but \cite{orusa+23} showed that it is realized at least for $M_{\rm A} \gtrsim 100$;
for $M_{\rm A} \lesssim 100$, instead, spectra are typically steeper.
Although injection at the FS can be adjusted to produce any spectral index, in this section we consider a shock with $M_{\rm A} = 100$ ($v_1 = -100 v_{\rm A}$ as seen from the downstream frame) and $\thbn=80^\circ$, and use the same CR injection as in Run $\mathcal{A}$ to produce a $p^{-4}$ spectrum (Run $\mathcal{B}$) to be compared with its RW homologous \citep[run E in][, here labeled as 3D in Table \ref{table:sim_parameters}]{orusa+23}.
Since the latter was performed with the strictly non-relativistic version of \texttt{dHybrid}, in the former we consider CRs with an initial isotropic momentum $p_{\rm iso} \ll mc$, so that they remain non-relativistic over the duration of our comparison ($\sim 115 \,\omega_{\rm c}^{-1}$).
\begin{figure*}
    \centering
    \subfigure[$B_{tot}$]{
        \includegraphics[trim = 30 40 40 15, clip=true, width=0.45\textwidth]{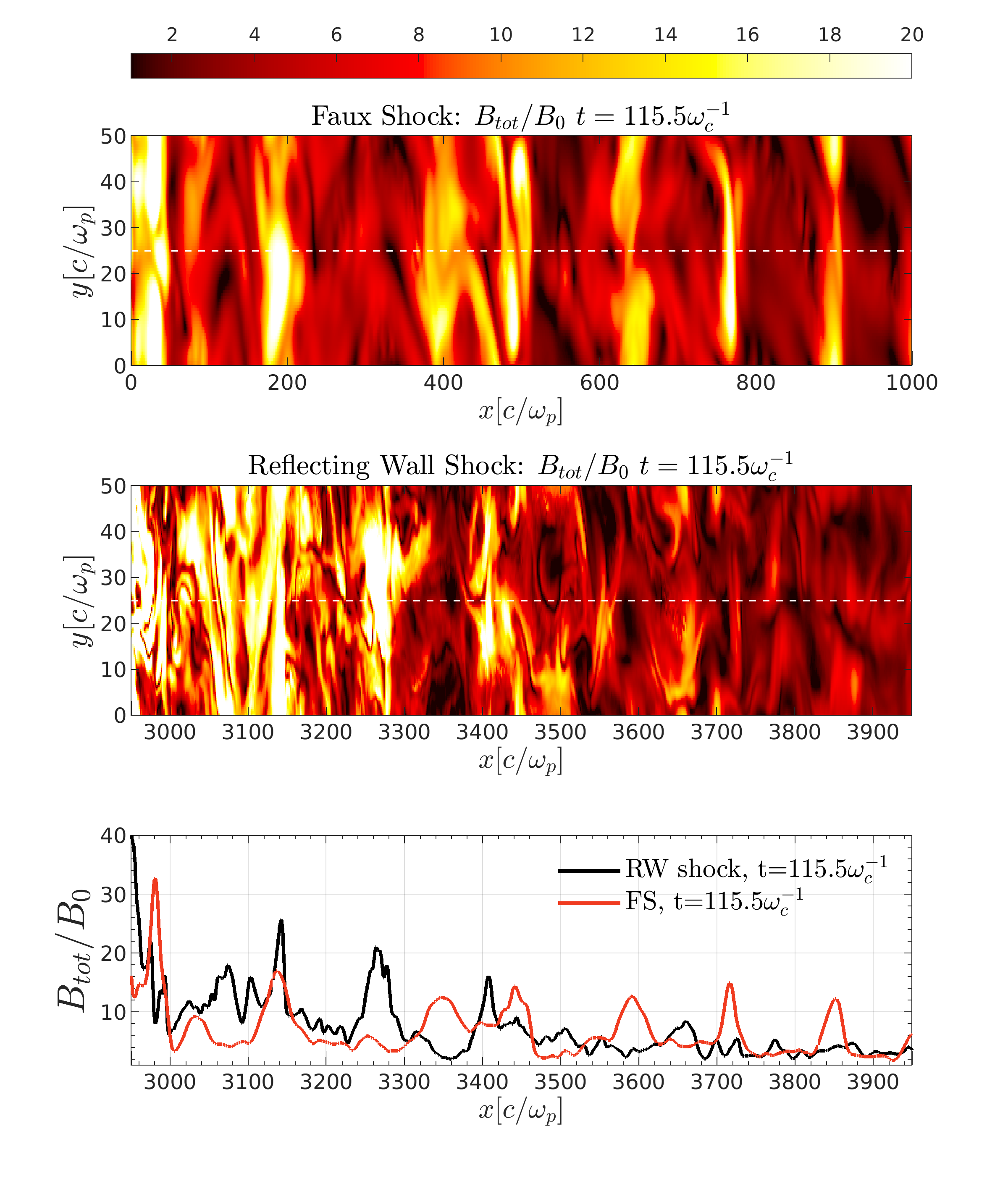}
    }
    \hspace{-0.3cm}
    \subfigure[$B_x$, $B_y$, $B_z$]{
        \includegraphics[trim = 20 15 55 40, clip=true, width=0.45\textwidth]{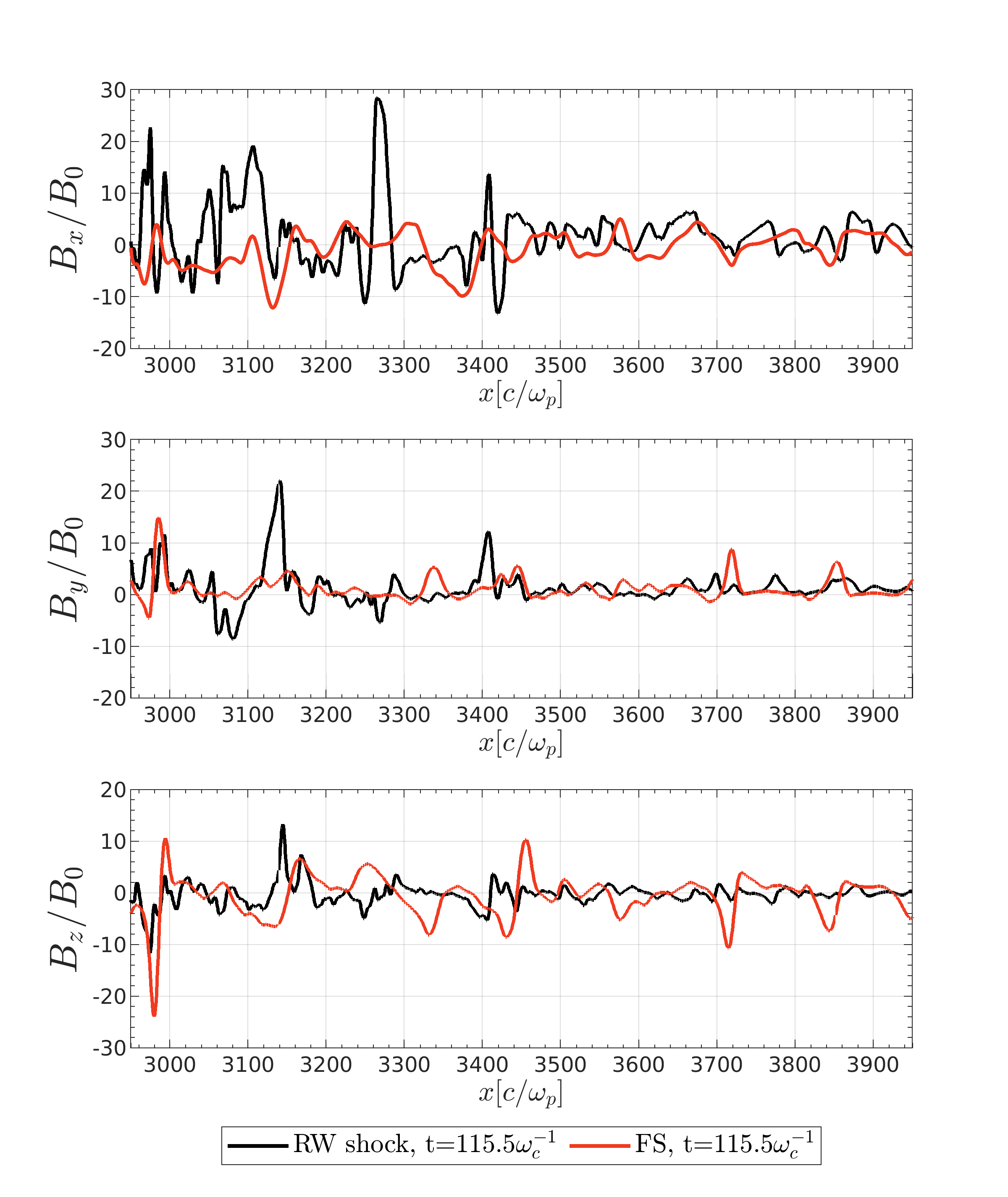}
    }
    \caption{(a) Magnetic field strength for the $80$ degree perpendicular FS, Run $\mathcal{B}$, (top panel) and the $80$ degree 3D RW shock, Run $3\mathcal{D}$, (middle panel) together with lineouts (taken from $y=150 \, d_{\rm i}$, dashed white lines) of the magnetic field strength (bottom panel) at $t = 115.5 \, \omega_{\rm c}^{-1}$. (b) Lineout comparison of each component of the magnetic field ($B_x$, $B_y$, and $B_z$) for the $80$ degree FS simulation (red) versus the $80$ degree RW shock simulation (black).}
    \label{fig:perp_Bfields}
\end{figure*}

\subsubsection{Momentum Spectra}
As before, Figure \ref{fig:momentum_RunB} shows the momentum phase space distributions and the FS/RW spectra both at the same timestep (solid lines) and the offset time $\Delta t = 43 \, \omega_c^{-1}$ (dotted lines).
The FS upstream length is truncated so that the visible upstream length is the same for both Runs.
The FS and $3\mathcal{D}$ RW shock show good agreement in their non-thermal phase space distributions, including having the same normalization, validating our prescription for particle injection even for oblique shocks.
Both Runs $\mathcal{B}$ and $3\mathcal{D}$ are initialized with 4 particles per cell, which leads to a larger overall density of CRs for the RW shock simulation which has a larger number of cells along all three dimensions.
However, the number density of CRs in the FS could be easily increased if necessary without approaching the computational cost of Run $3\mathcal{D}$.

Figure \ref{fig:momentum_RunB}(a) shows the total momentum phase space, and the FS reproduces the flat non-thermal spectrum found in the RW shock at the offset time. 
The bottom panel compares momentum spectra.

Subfigure \ref{fig:momentum_RunB}(c) shows a population of CRs that appear confined close to the shock with both positive and negative values of $p_y$, as well as a population streaming away upstream with only $p_y>0$ values;
since $\mathbf{y}$ is essentially the direction of $\mathbf{B}_0$, this is essentially the analog of the $p_x$ phase space for the parallel shock in Figure \ref{fig:momentum_RunA}.
Also the transverse phase spaces (i.e., in $p_x$ and $p_z$) are remarkably similar between the FS and RW setups, as shown in subfigures \ref{fig:momentum_RunB}(b) and (d). 
The $3\mathcal{D}$ RW shock has visible boundary effects at the far right edge of the box in both $p_x$ and $p_z$ due to the left-handedness of CR proton gyration.
These effects are also visible in the FS at its far right edge, but are not depicted in Figure \ref{fig:momentum_RunB}, which is truncated $3000  \, d_{\rm i}$ before the end of the box.

In general, being able to run oblique shocks in 2D, though with a FS injector, unlocks the possibility of studying the long-term evolution of this systems without numerical artifacts, which is effectively precluded in full 3D RW setups. 
A detailed study of particle acceleration at perpendicular shocks is left to a future work. 

The spectra (bottom panels in Figure \ref{fig:momentum_RunB}), even when compared using the best matching $\Delta t$ time offset, show minor but interesting deviations between the FW and RW setups.
In $n(p_x)$ and $n(p_y)$ the FS produces a harder spectrum for negative momenta, whereas in $n(p_z)$ it produces a harder spectrum for $p_z>0$.
We argue that the reason lies in our assuming that downstream scattering produces an isotropic CR distribution; 
it is likely that highly-oblique shocks do not fill the CR loss cone isotropically, thus affecting the average energy gain and return probability. 
These modifications manifest themselves in spectral deviations, which are observed for oblique shocks with $M_A\lesssim 100$ \citep{orusa+23}.
Yet, the FS setup has the flexibility to adapt and reproduce these effects as we refine our understanding of injection with traditional RW shock simulations.

\subsubsection{Magnetic Fields} 
Subfigure \ref{fig:perp_Bfields}(a) shows the magnetic field strength over $1000 \, d_{\rm i}$ ahead of the shock at the final timestep in Run $\mathcal{B}$ (top panel), Run $3\mathcal{D}$ (middle panel), and a horizontal slice of the total magnetic strength of both Runs at $t = 115.5 \, \omega_{\rm c}^{-1}$(bottom panel). 
Because of the computational cost of Run $3\mathcal{D}$, as well as its diminishing upstream length as a function of time, we have relatively few cyclotron times available with which to compare to the FS in Run $\mathcal{B}$, which itself requires many cyclotron times to converge into a self-consistent upstream behavior after the initial transient has passed.
The result is that we must neglect early times of the FS, and we cannot access late times of the $3\mathcal{D}$ RW shock.
Therefore, we compare only the same simulation timestep without showing an equivalent offset time between the two.
Bearing this offset in mind, the amplitude of the magnetic field strength and the approximate wavelength between peaks are fairly consistent in both runs.

However, like in the parallel case, there are some clear qualitative differences.
First, the presence of large amplitude fluctuations extends further upstream in the FS than in the RW, which can be attributed to the time offset and the greater upstream extent in Run $\mathcal{B}$ at this timestep.
Second, and more interesting, is the presence of smaller scale modes in the 3$\mathcal{D}$ RW shock.
Besides the general comment that magnetic turbulence works differently in 2D and 3D, we do not have a clear explanation for what causes these differences;
some effects of magnetic field growth/evolution dependence on simulation dimensionality have been discussed in the literature \citep{trotta+19, boula+24, zacharegkas+24}, and it is possible that the lack of small scale-modes could be an unavoidable consequence of reduced dimensionality.

Beyond this, large Mach number shocks in RW simulations can develop a complex spatial structure at the shock foot spanning many tens of $d_{\rm i}$ along $x$ for oblique cases and even thousands of $d_{\rm i}$ in quasi-parallel ones \citep{caprioli+14a,caprioli+14b}.
This makes a singular shock location hard to precisely define, whereas the FS is defined to be a perfectly smooth boundary at $x=0$.
Therefore aligning the positions of the shock in FS and RW high-$M_A$ shocks is imprecise, and this may also contribute to differences in magnetic field strength and extent in the upstream. 

Subfigure \ref{fig:perp_Bfields}(b) shows the individual components of the magnetic field for Runs $\mathcal{B}$ and 3$\mathcal{D}$ averaged over the $y-$dimension.
As is mentioned in the Appendix of \cite{zacharegkas+24}, the self-generated magnetic field components perpendicular to $\mathbf{B}_0$ in 2D hybrid simulations with currents aligned with $\mathbf{B}_0$ tend to be anisotropic, with the out-of-plane component typically stronger than the in-plane component. This out-of-plane field more closely resembles the structure found in full 3D simulations.
It has not yet been studied how this effect carries over to highly-oblique shocks, but we qualitatively note that when  $\mathbf{B}_0$ is directed almost entirely along the in-plane $y-$axis, the out-of-plane $B_z$ component of both Runs B and D seem to be in reasonably good agreement, whereas the FS has significantly lower amplitude in the other in-plane quasi-perpendicular component, $B_x$. 
$B_y$ is also noticeably discrepant between Runs $\mathcal{B}$ and $3\mathcal{D}$, which may be a byproduct of more turbulence from small-scale modes in Run $3\mathcal{D}$. 

Despite these caveats and discrepancies, especially at small scales, the close agreement in total magnetic fields (Figure \ref{fig:perp_Bfields}a) and maximum CR energies (Figure \ref{fig:momentum_RunB}) lends credence to the idea that the relevant physics is still being captured with the FS setup.

\section{Discussion \& Conclusions}
\label{sec:conclusions}

We introduced a simplified computational scheme for studying CR acceleration at non-relativistic shocks, which hinges in introducing a \textit{faux shock} (FS) boundary condition that injects a prescribed population of CRs, which then evolves self-consistently under the action of self-generated magnetic fluctuations. 

We demonstrated that the FS setup successfully captures most of the CR physics found in traditional simulations with a piston or a reflective wall (RW) both at parallel and quasi-perpendicular shocks.
The importance of this novel simulation setup is that it renders the study of self-driven CR currents in far upstream feasible for large boxes, high Mach numbers, and different shock obliquities, enabling us to study plasma instabilities (such as the Bell instability) over a wide range of parameters with direct implications for CR acceleration at astrophysical sources.

The limitations to this setup are that it is not \textit{fully} self-consistent in the sense that the shock is represented by a sharp discontinuity (a complex turbulent transition is expected at large $M_A$ shocks) and that CR injection is prescribed rather than self-consistently injected from the thermal pool.
Current global hybrid simulations can be used to calibrate the flux and isotropy of the injected CRs, and are leveraged in this work to push forward a comparison between FS and RW computational frameworks. 

After a few tens of cyclotron times in which the FS simulation is dominated by a spurious CR transient (which is present also in RW simulations, though for a shorter amount of time), both particle spectra and electro-magnetic structures in the precursor evolve similarly and are consistent in both setups.

While for quasi-parallel shocks \S\ref{ssec:parallel} this could be expected, it is non trivial to find a good agreement for quasi-perpendicular shocks \S\ref{ssec:oblique}, which need to be modeled in 3D in a RW setup to capture ion injection accurately \citep{jones+98, orusa+23}, but are here modeled in 2D with a FS scheme.

In such oblique cases, we report small differences in anisotropy of the CR populations and in the size and structure of magnetic turbulence between the FS and RW shock setups, the exact cause of which is not fully understood.
Yet even with these differences, FS spectra replicate the RW shock simulations closely, indicating that the differences in the magnetic field are not critical to the overall picture of CR acceleration, nor to the growth of CR energies in time.

When used in conjunction with traditional RW simulations that establish the appropriate energy gain and escape probability per acceleration cycle, as well as the normalization and the fraction of injected particles, the FS setup can be a useful tool to explore regimes that are otherwise computationally inaccessible.
It is highly flexible such that any nonrelativistic shock speed and shock obliquity can be simulated by adjusting the FS boundary condition, which could be made time-dependent if the shock dynamics evolved with time, e.g., because CR acceleration changes with time \citep{haggerty+20, caprioli+20}.

Beyond this, it allows for a dramatic increase in the statistics of CRs, lending itself to a deeper study of CR dynamics like the spatial- and momentum-dependence of the CR diffusion coefficient and the structures within CR phase spaces. 
The application of this tool will be used to investigate the maximum energy CRs produced at supernova remnants in forthcoming works. 

\section{Acknowledgments}
The authors thank Luca Orusa for providing the 3D perpendicular simulation (Run $3\mathcal{D}$) for our comparison analysis. 
E.S.~was partially supported by the NSF Graduate Research Fellowship Program (grant 2140001).
D.C.~was partially supported by NASA (grants 80NSSC24K0173, S02028-01, and 80NSSC23K1481) and NSF (grants PHY-2010240 and AST-2308021).
C.C.H.~was partially supported by NSF (grants AGS-1936393 and AGS-2338131) and NASA (grants 80NSSC24K0173 and 80NSSC23K0099).

\bibliography{Total}{}
\bibliographystyle{aasjournal}



\end{document}